\documentclass[aps,pra,reprint,groupedaddress,superscriptaddress]{revtex4-1}

\usepackage{graphicx}
\usepackage{amsmath}

\begin{document}


\title{Entanglement of remote material qubits through nonexciting interaction with single photons}


\author{Gang Li}
\email[]{gangli@sxu.edu.cn}
\author{Pengfei Zhang}
\author{Tiancai Zhang}
\email[]{tczhang@sxu.edu.cn}
\affiliation{State Key Laboratory of Quantum Optics and Quantum Optics Devices,
and Institute of Opto-Electronics, Shanxi University, Taiyuan 030006, China}
\affiliation{Collaborative Innovation Center of Extreme Optics, Shanxi University, Taiyuan 030006, China}


\date{\today}

\begin{abstract}
We propose a scheme to entangle multiple material qubits through interaction with single photons via non-exciting processes associated with strongly coupling systems. The basic idea is based on the material state dependent reflection and transmission for the input photons. Thus, the material qubits in several systems can be entangled when one photon interacts with each system in cascade and the photon paths are mixed by the photon detection. The character of non-exciting of the material qubits does not change the state of material qubit and thus ensures the possibility of purifying entangled states by using more photons under realistic imperfect parameters. It also guarantees directly scaling up the scheme to entangle more qubits. Detailed analysis of fidelity and success probability of the scheme in the frame of an optical Fabry-P\'{e}rot (FP) cavity based strongly coupling system is presented. It is shown that a two-qubit entangled state with fidelity above 0.99 is promised with only two photons by using currently feasible experimental parameters. Our scheme can also be directly implemented on other strongly coupled system.
\end{abstract}



\maketitle

\section{Introduction}
Quantum entanglement is a one of the key features in quantum mechanics and has been recognized as an important resource for quantum information processing \cite{Nielsen00} and quantum measurement \cite{Roos06}. Entanglement of remote material qubits is the essential intergradient for long distance quantum communication \cite{Briegel98, Duan01, Sangouard11} and quantum networks \cite{Kimble08}. There have been several methods used to produce remote entanglement among qubits and all of these methods involve the excitation of the material qubits and the process of emission or absorption of photons. The first technique involves entangling a photon to the first material qubit and directly writing it into the second material qubit \cite{Ritter12, Matsukevich06}. The second one is a heralded protocol \cite{Duan01, Duan03}, wherein two photons entangled with each of the two material qubits interfere in a 50/50 beam splitter. Upon different combination of detected photon states, the states of material qubits are projected into various entangled states \cite{Chou05, Hofmann12, Lee11, Yuan08, Moehring07, Chou07}. The third way is based on the quantum interference of two separated atomic qubits \cite{Cabrillo99}. The detection of a single photon from two atomic emissions produces entanglement between them \cite{Slodi13}.

In a strong coupling system, a single material particle can interact with photons without exciting the material particle. In this nonexciting regime, the path of the input photon is determined by the states of the material particle. In state where the particle strongly couples to optical cavities \cite{Volz11, Aoki09, Shomroni14, Scheucher2016} or other structures, such as the nanoscale surface plasmons \cite{Chang07}, the incident photon will be reflected. In the state where the particles do not couple, the photon will transmit. With such systems nondestructive detection of atoms \cite{Volz11} and quantum single photon circulators controlled by the state of the single atom have been experimentally demonstrated \cite{Scheucher2016}. A single-photon transistor based on nanoscale surface plasmons has also been theoretically proposed \cite{Chang07}.

Here we propose a scheme to produce remote entanglement based on this non-exciting interaction between the atom and photon in a strongly coupling system. The nonexciting process does not change the state of the material qubit and thus ensures the possibility of purifying the entangled state by using more photons in a real situation where loss and other imperfections are inevitable. This also guarantees scalability for producing entanglement among additional qubits. So, our proposal can be easily scaled up to create entanglement among multiple nodes in a quantum network \cite{Kimble08}, or to generate the Greenberger-Horne-Zeilinger (GHZ) state among remote material qubits \cite{Greenberger09}.

We first introduce the basic ideas in section II, where the theories of state-dependent reflection and transmission in a strongly coupling system, entangling two qubits in two systems by one photon, and scaling up the scheme are presented. Next, detailed analysis of state fidelity and success probability of our scheme in the frame of an optical Fabry-P\'{e}rot (FP) cavity based strongly coupling system is given in section III, in which a set of Heisenberg equations is built to simulate the reflection and transmission of the FP cavity with pulsed input single photons. At last, we conclude our paper and discuss the outlook for realization of our scheme on possible experiment systems in section IV.

\section{Basic idea}
\begin{figure}
\includegraphics[width=8.5cm]{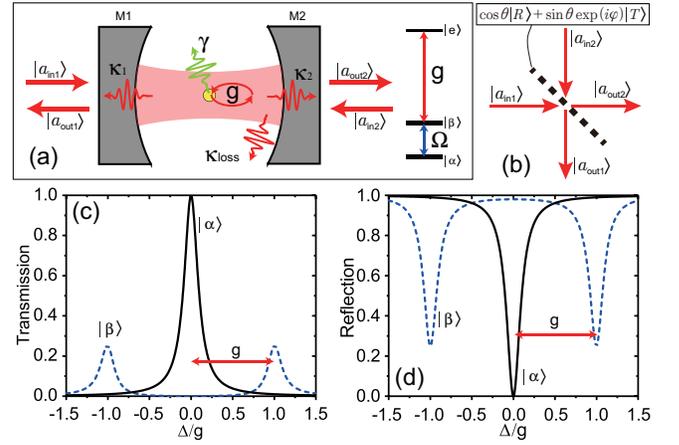}
\caption{\label{fig1} (Color online) (a) Schematic of a strongly-coupled CQED system with a single atom residing in the FP cavity. (b) A system equivalent to (a) with a four-port quantum router with internal states represented by $|R\rangle$ and $|T\rangle$. The power transmission (c) and reflection (d) spectra for a cavity with atoms in states $|\alpha\rangle$ and $|\beta\rangle$ are calculated by using parameters $g=10\kappa$, $\gamma=\kappa$, $\kappa_1=\kappa_2=\kappa/2$, and $\Delta_\text{a}=\Delta_\text{c}=0$.}
\end{figure}

We first take a strongly coupling CQED system between single atoms and an optical FP cavity \cite{Boca04, Maunz05} as an example to describe the state-dependent reflection and transmission of the incident photons. Other systems, such as strongly coupling surface-plasmon-emitter systems and strongly coupling systems between single atom and whispering-gallery-mode optical microresonators, work in the same way. The basic concept is shown in Fig.~\ref{fig1}(a), where the cavity is comprised of two mirrors with transmission decay rates of $\kappa_{1(2)}$, and $\kappa_\text{loss}$ represents the decay rate of unexpected cavity loss from the scattering and absorption. An atom with two ground states, $|\alpha\rangle$ and $|\beta\rangle$, and an excited state $|e\rangle$ associated with decay rate $\gamma$, resides in the cavity. The atomic transition $|\beta\rangle \leftrightarrow |e\rangle$ strongly couples to the cavity mode with coupling strength $g$. The Hamiltonian of the system is given by
\begin{equation}\label{eq1}
  H=\hbar g \left(|\beta\rangle \langle e|a^{\dagger}+|e\rangle \langle \beta|a\right)
\end{equation}
with $a$ and $a^{\dagger}$ being the annihilation and creation operators of the cavity mode. If a weak coherent light beam $a_{in1(2)}$ is incident on mirror M1(2), the dynamics of the intracavity field $a$ is then described by the Heisenberg-Langevin equation \cite{Walls2008, Duan04},
\begin{equation}\label{eq2}
  \dot{a}=-\frac{i}{\hbar}[a,H]-\kappa a + \sqrt{2 \kappa_{1(2)}} a_\text{in1(2)},
\end{equation}
where $a$ is the field amplitude of cavity mode. At the same time, we have the relations between the input and output fields of the cavity
\begin{equation} \label{eq3}
  a_\text{out1}+a_\text{in1}=\sqrt{2\kappa_1} a
\end{equation}
and
\begin{equation}\label{eq4}
  a_\text{out2}+a_\text{in2}=\sqrt{2\kappa_2} a,
\end{equation}
where $a_\text{out1}$ and $a_\text{out2}$ are the cavity output fields from M1 and M2. Under weak excitation approximations, which means the excitation of the atom is negligible, Eqs.~(\ref{eq2}) and (\ref{eq3}) can be analytically solved and the coefficients for reflection and transmission of input field $a_\text{in1(2)}$ are
\begin{equation} \label{eq5}
  r_{1(2)}=\frac{a_\text{out1(2)}}{a_\text{in1(2)}}=1-\frac{2 \kappa_{1(2)} (i \Delta_\text{a} +\gamma)}{(i\Delta_\text{c}+\kappa)(i\Delta_\text{a}+\gamma)+g^2}
\end{equation}
and
\begin{equation}\label{eq6}
  t_{1(2)}=\frac{a_\text{out2(1)}}{a_\text{in1(2)}}=\frac{2 \sqrt{\kappa_{1} \kappa_{2}} (i \Delta_\text{a} +\gamma)}{(i\Delta_\text{c}+\kappa)(i\Delta_\text{a}+\gamma)+g^2},
\end{equation}
here $\Delta_\text{a}$ and $\Delta_\text{c}$ are the frequency detunings of the incident field with respect to the atomic transition and cavity. In our case $\Delta_\text{a}=\Delta_\text{c}=0$, the reflectivity and transmission can be simplified as
\begin{equation}\label{eq7}
  r_{1(2)}=1-2\kappa_{1(2)} \gamma/(\kappa \gamma+g^2)
\end{equation}
and
\begin{equation}\label{eq8}
  t_{1(2)}=2 \sqrt{\kappa_1 \kappa_2} \gamma/(\kappa \gamma+g^2).
\end{equation}
We can see the coefficients of transmission from the both sides are the same. In the ideal case, where $\kappa_{\text{loss}}=0$ and $\kappa_1=\kappa_2$ the reflectivity and transmission are
\begin{equation}\label{eq9}
  r=1-\kappa \gamma/(\kappa \gamma+g^2)
\end{equation}
and
\begin{equation}\label{eq10}
t=\kappa \gamma/(\kappa \gamma+g^2).
\end{equation}

If the atom is in state $|\alpha\rangle$, the coupling efficiency $g=0$, thus we get $r=0$ and $t=1$ as the behavior of an empty cavity, see Fig.~\ref{fig1}(c) and (d). The photon will transmit through the cavity without interacting with the atom. If the atom is in state $|\beta\rangle$, where the atom-cavity coupling $g$ is switched on, due to the normal mode splitting $2g$ the incident photon becomes detuned from the coupled states and thus is reflected. We will get $r\approx 1$ and $t\approx 0$ if $g \gg \kappa, \gamma$ from Eqs. ~(\ref{eq9}) and (\ref{eq10}). In both of these cases, there is no excitation of the atom and, in principle, the atomic state will not be changed. Thus by setting the atomic state in $|\alpha\rangle$ or $|\beta\rangle$, the CQED system can route the incident photon from the input mode $|a_\text{in1}\rangle$ ($|a_\text{in2}\rangle$) to output modes $|a_\text{out2}\rangle$ ($|a_\text{out1}\rangle$) or $|a_\text{out1}\rangle$ ($|a_\text{out2}\rangle$), respectively. If the atom is in a superposition state $\cos \theta |\alpha\rangle + \sin \theta \exp{i\varphi}|\beta\rangle$ the system will route the input photons to both of the output modes with probabilities $\cos^2 \theta$ and $\sin^2 \theta$. As such, the system works exactly as a single-photon quantum router.

This system is equivalent to a four-port quantum optical beam splitter (or circulator) where the reflection and transmission of incident photons are controlled by the state of the strongly coupled atom. The corresponding schematic is shown in Fig.~\ref{fig1}(b), where we use $|R\rangle$ and $|T\rangle$ as meaningful representations of the internal states of the strongly coupled material qubit. If the state of a router is $|R\rangle$, photons from both input modes would be reflected. In other words, the input photons in $|a_\text{in1}\rangle$ and $|a_\text{in2}\rangle$ are routed to $|a_\text{out1}\rangle$ and $|a_\text{out2}\rangle$, respectively. Otherwise, the $|T\rangle$-state router will route photons from $|a_\text{in1}\rangle$ to $|a_\text{out2}\rangle$ and photons from $|a_\text{in2}\rangle$ to $|a_\text{out1}\rangle$ via transmission.

\begin{figure}
\includegraphics[width=8.5cm]{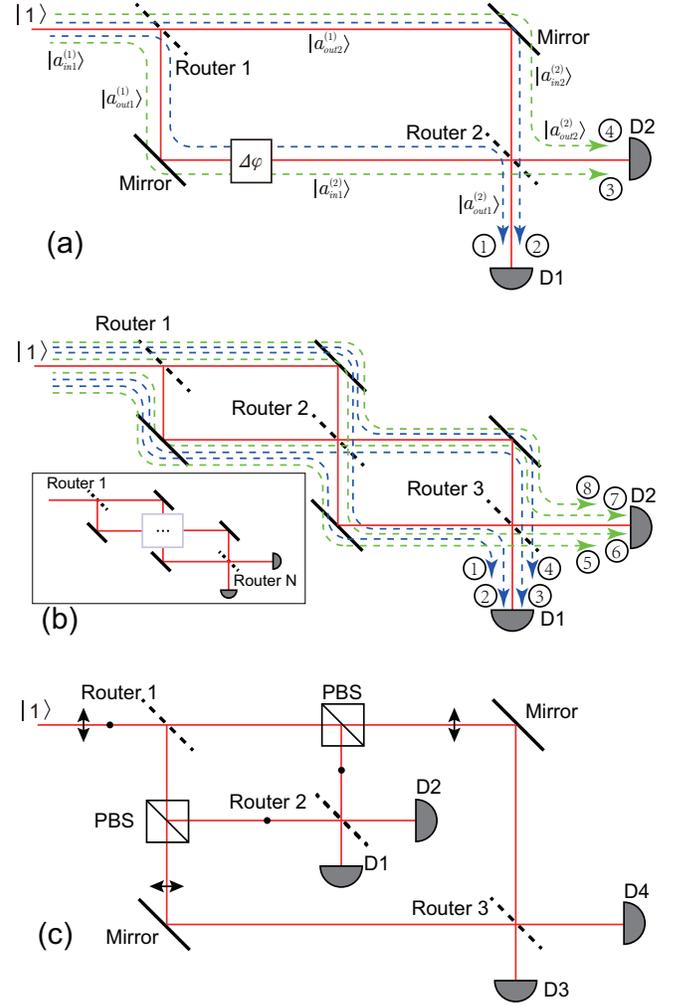}
\caption{\label{fig2} (Color online) Schematics of producing entangled states between two quantum routers (a), 3 router-entangled state (b) and GHZ state among the 3 routers (c). The inset in (b) shows one method to scale up our scheme and generate N-router entangled state. In (b) and (c) the phase difference between different paths are taken as 0 for simplicity. In all the figures, each router has been prepared in an internal quantum state $[|R_n\rangle+\exp (i\varphi_n)|T_n\rangle]/\sqrt{2}$, in which $n$ means the router number.}
\end{figure}

Quantum entanglement of two material qubits can be realized using the configuration shown in Fig.~\ref{fig2} (a). Here two quantum routers are arranged to replace the two classical beam splitter in a Mach-Zehnder interferometer. Two single photon detectors, D1 and D2, are used to detect photons from the two output modes $|a_\text{out1}\rangle$  and $|a_\text{out2}\rangle$. Quantum states of the routers are initially prepared in its maximum coherent superposition, i.e., $[|R_1\rangle+\exp (i\varphi_1)|T_1\rangle]/\sqrt{2}$ and $[|R_2\rangle+\exp (i\varphi_2)|T_2\rangle]/\sqrt{2}$. The overall quantum state is the direct product of these two wave functions. By sending a single photon into the input mode $|a_{\text{in}1}^{(1)}\rangle $, the overall state for the whole system after the photon is transmitted through the two cascade systems is expressed as:

\begin{equation}\label{eq11}
\begin{aligned}
|\Psi_{\text{out}}\rangle = &\frac{1}{2}\left[|R_1,R_2\rangle+ e^{i (\varphi_1+\varphi_2 +\Delta \varphi)}|T_1,T_2\rangle\right]|a^{(2)}_{\text{out}1}\rangle \\
&+\frac{1}{2}\left[e^{i (\varphi_2 +\Delta \varphi)}|R_1,T_2\rangle+e^{i \varphi_1}|T_1,R_2\rangle\right]|a^{(2)}_{\text{out}2}\rangle \\
= &|\Phi_2\rangle|a^{(2)}_{\text{out}1}\rangle/\sqrt{2}+ |\Psi_2\rangle|a^{(2)}_{\text{out}2}\rangle/\sqrt{2}
\end{aligned}
\end{equation}
with
\begin{equation}\label{eq12}
|\Phi_2\rangle=\left[|R_1,R_2\rangle+e^{i (\varphi_1+\varphi_2 + \Delta \varphi)}|T_1,T_2\rangle\right]/\sqrt{2}
\end{equation}
and
\begin{equation}\label{eq13}
|\Psi_2\rangle=\left[e^{i (\varphi_2 +\Delta \varphi)}|R_1,T_2\rangle+e^{i \varphi_1}|T_1,R_2\rangle\right]/\sqrt{2}
\end{equation}
exactly representing the two maximum entangled states for the routers. Here $\Delta \varphi$ is the phase difference between two paths of the interferometer. From Eq.~(\ref{eq11}) we can see that upon the event of photon detection by D1 or D2 the overall state of the two routers collapses into $|{{\Phi }_{2}}\rangle $ or $|{{\Psi }_{2}}\rangle $, respectively. The process of entanglement can be understood as path 1 and path 2 (3 or 4) which can not be distinguished from each other by photon detection with D1 (D2), as shown in FIG. 2. As such, a photon click in D1 (D2) will lead to entanglement showing in Eq.~(\ref{eq12}) [Eq.~(\ref{eq13})]. The probability of detecting a single photon by either of the two detectors is 0.5, which implies the probability of preparing each maximum entangled state is 50\%. It should be emphasized that local operations on each material qubit, such as ground state operation of a single atom either by the two-photon Raman process \cite{Wang2014} or microwave driving \cite{Xia15}, can be applied separately. Thus, upon the click of D2 $|{{\Psi }_{2}}\rangle $ can be converted into $|{{\Phi }_{2}}\rangle $ by applying a local $\pi$ rotation and a phase shift on router 2 and vice versa. Thus, in theory, entangled state $|{{\Phi }_{2}}\rangle $ or $|{{\Psi }_{2}}\rangle $ can be prepared with probability of 1 on demand.

Because the material qubits do not absorb the incident photon and states of the material qubits remains unchanged after the photon has been detected, our scheme can be directly scaled up to entangle more material qubits. Figs.~\ref{fig2} (b) and (c) show two configurations that can be used to produce different types of entangled states among three or more qubits. In Fig.~\ref{fig2}(b), a cascade configuration of three quantum routers is displayed, where two photon detectors are used. The three involved qubits, initially prepared in their maximum superposition state $|R_n\rangle+\exp{i\phi_n |T_n\rangle}$ ($n=1,2,3$), can be entangled by sending and detecting single photon. When D1 or D2 clicks, the entangled state of
\begin{equation}\label{eq14}
|\Phi_3\rangle=\left(|R_3\rangle|\Phi_2\rangle+e^{i \varphi_3}|T_3\rangle|\Psi_2\rangle\right)/\sqrt{2}
\end{equation}
or
\begin{equation}\label{eq15}
|\Psi_3\rangle=\left(|R_3\rangle|\Psi_2\rangle+e^{i \varphi_3}|T_3\rangle|\Phi_2\rangle\right)/\sqrt{2}
\end{equation}
is prepared with $|\Phi_2\rangle$ and $|\Psi_2\rangle$ given by Eqs.~(\ref{eq12}) and~(\ref{eq13}). This can also be explained by the fact that a click on D1 (D2) cannot distinguish photon paths among 1--4 (5--8), as shown in FIG.~\ref{fig2}(b), and then entangles the internal states of routers. This cascade configuration can be directly scaled up as the schematic in the inset of Fig.~\ref{fig2}(b) to produce large scale entangled states among $N$ quantum routers.

By following a similar process, a GHZ state with three qubits can be generated if we take the polarization of photons into account. As shown in Fig.~\ref{fig2}(c), detectors D1 and D2 (D3 and D4) are arranged to detect photons with vertical (horizontal) polarization. Two polarization beam splitters (PBSs) are used to connect router 1 to router 2 (3) by photons with vertical (horizontal) polarization. After a click of D1 or D2 and a local operation on a qubit in router 2, a maximum entangled state [Eq.~(\ref{eq12})] is produced between routers 1 and 2. Then by using a single photon with horizontal polarization as the input, a click of D3 or D4 and a corresponding local operation on qubits in router 3 will produce a maximum entangled state $[|R_1,R_3\rangle+\exp{i (\varphi_1+\varphi_3)}|T_1,T_3\rangle]/\sqrt{2}$ between router 1 and router 3. The overall state is obvious a GHZ state $|\text{GHZ}\rangle =[|{{R}_{1}},{{R}_{2}},{{R}_{3}}\rangle +\exp i({{\varphi }_{1}}+{{\varphi }_{2}}+{{\varphi }_{3}})|{{T}_{1}},{{T}_{2}},{{T}_{3}}\rangle]/\sqrt{2}$. Here we omit the phase difference between the photon paths for simplicity.

\section{Implementation of the scheme in a strongly coupling system with single atoms and an optical FP cavity}
We now consider experimental realization of our scheme in a strongly coupled system with single atoms coupling to an optical FP cavity. We will show the achievable state fidelity and the success probability of our scheme in a situation where experimental imperfections, like limited coupling strength $g$, slow response of the cavity and extra cavity losses, are taken into account.

In order to simulate the response of the CQED system to a single photon pulse incident on one of the cavity mirrors we assume that all the input photon pulses follow a Gaussian shape $f_\text{in1(2)}(t) = C_N \exp{[-(t-T/5)^2/T^2]}$, where $C_N$ is the normalizing factor, $T$ is the total pulse length and $t$ ranges from 0 to $T$. So we have $\int_0^{T} |f_\text{in1(2)}(t)|^2=1$. Thus, the input single photons have the form $|\psi_\text{in1(2)}(t)\rangle=\int_0^{T} f_\text{in1(2)}(t) a_\text{in1(2)}^\dagger(t) dt |\text{vac}\rangle$ with $[a_\text{in1(2)} (t),a_\text{in1(2)}^\dagger (t') ]=\delta(t-t')$ and $|\text{vac}\rangle$ representing the vacuum state. Similarly, we can also define the output single photon pulse by $|\psi_\text{out1(2)}(t)\rangle=\int_0^{T} f_\text{out1(2)}(t) a_\text{out1(2)}^\dagger (t) dt |\text{vac}\rangle$ with similar commutation relation $[a_\text{out1(2)}(t),a_\text{out1(2)}^\dagger (t') ]=\delta(t-t')$ and $f_\text{out1(2)}(t)$ the output pulse shape. Since the input pulse shape is given we can get the output pulse shape through standard Heisenberg-Langevin equations.

When the atom is in state $|\beta\rangle$ the atom is strongly coupling to the cavity mode. The time evolution of cavity field $a$ with single-photon pulses incident on both sides of the cavity is then \cite{Walls2008}
\begin{equation}\label{eq16}
  \dot{a}=-\frac{i}{\hbar}[a,H']-\kappa a - \sum\limits_{j=1,2}\sqrt{2 \kappa_j} a_{\text{in}j},
\end{equation}
where the Hamiltonian is
\begin{equation} \label{eq17}
  H'=H-i\gamma \sigma_{ee},
\end{equation}
with $\sigma_{ee}=|e\rangle \langle e|$. The relations between input and output fields are given by Eqs. (\ref{eq3}) and (\ref{eq4}). The Heisenberg equation for the atomic operators is
\begin{equation} \label{eq18}
  \dot{\sigma} = -i [\sigma, H'],
\end{equation}
with $\sigma=|\beta\rangle \langle e|$. Thus by using Eqs.~(\ref{eq16}) and (\ref{eq18}) we get two operator equations which describe the dynamics of the CQED system when the single-photon pulse is incident on one side of the cavity. They are
\begin{equation} \label{eq19}
 \begin{split}
  \dot{a} = & -i g \sigma - \kappa a - \sum\limits_{j=1,2}\sqrt{2 \kappa_j} a_{\text{in}j}, \\
  \dot{\sigma} =  & - i g a (\sigma_{ee}-\sigma_{\beta \beta}) - \gamma \sigma.
 \end{split}
\end{equation}
These equations together with Eqs.~(\ref{eq2}) and (\ref{eq3}) describe the whole dynamics of the CQED system.

Next we will transform these equations from the Heisenberg picture to the Shr\"odinger picture and finally solve the reflected and transmitted pulse shapes. Since we are only considering one-photon excitation, the time-dependent wave function of the system can be defined as
\begin{equation} \label{eq20}
\begin{split}
  |\psi(t)\rangle = & C_\beta(t)|\beta, 1, \text{vac1}, \text{vac2} \rangle + C_e(t)|e, 0, \text{vac1}, \text{vac2} \rangle \\
  & + \sum\limits_{j=1,2} \int_0^T [ f_{\text{in}j}(t) a_{\text{in}j}^\dagger(t) + f_{\text{out}j}(t) a_{\text{out}j}^\dagger(t)] dt \\
  & \times |\beta, 0, \text{vac1}, \text{vac2}\rangle,
\end{split}
\end{equation}
where $|x, 1, \text{vac1}, \text{vac2}\rangle$ denotes the atom is in state $|x\rangle$ ($x$ can be $\beta$ or $e$), the cavity contains one photon, and both of the fields outside of cavity are in the vacuum state; $C_x(t)$ is the corresponding time-dependent coefficient. Thus we have a set of equations in the Shr\"odinger picture (see the Appendix)
\begin{equation} \label{eq21}
 \begin{split}
  \dot{C}_{\beta}(t) = & -i g C_{e}(t) - \kappa C_\beta(t) - \sum\limits_{j=1,2}\sqrt{2 \kappa_j} f_{\text{in}j}(t), \\
  \dot{C}_{e}(t) = & - i g C_\beta(t) - \gamma C_{e}(t), \\
  f_\text{out1}(t) = & f_\text{in1}(t) + \sqrt{2\kappa_\text{1}} C_\beta(t),\\
  f_\text{out2}(t) = & f_\text{in2}(t) + \sqrt{2\kappa_\text{2}} C_\beta(t),
 \end{split}
\end{equation}
which describe the time evolution of time dependent coefficients and input and output pulse shapes.

When the atom is in state $|\alpha\rangle$ the atom does not couple to the cavity mode and the input single photon pulse encounters an empty cavity. In this case no atomic sate is involved and Eqs. (\ref{eq21}) become
\begin{equation} \label{eq22}
 \begin{split}
  \dot{C}(t) = & - \kappa C(t) - \sum\limits_{j=1,2}\sqrt{2 \kappa_j} f_{\text{in}j}(t), \\
  f_\text{out1}(t) = & f_\text{in1}(t) + \sqrt{2\kappa_\text{1}} C(t),\\
  f_\text{out2}(t) = & f_\text{in2}(t) + \sqrt{2\kappa_\text{2}} C(t),
 \end{split}
\end{equation}
where $C(t)$ is the coefficient for the state $|1, \text{vac1}, \text{vac2}\rangle$ with the cavity mode having one photon, and the two outside fields are in vacuum.

\begin{figure}
\includegraphics[width=8.5cm]{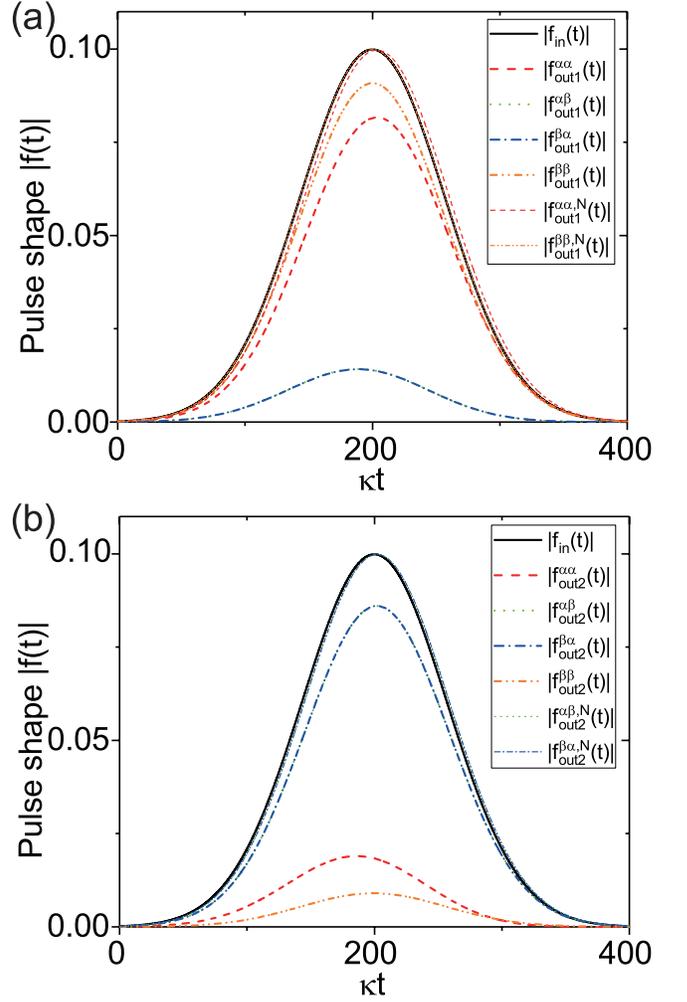}
\caption{\label{fig3} (Color online) Output pulse shapes from two output ports $|a^{(2)}_\text{out1}\rangle$ (a) and $|a^{(2)}_\text{out2}\rangle$ (b) with single a Gaussian pulse incident on $|a^{(1)}_\text{in1}\rangle$ port of interferometer in FIG.~\ref{fig2}(a) under different state combinations of two atoms. $|f_\text{in}(t)|$ is the amplitude variance of the input pulse. $|f^{xy}_\text{out1(2)}(t)|$ means the unnormalized output pulse amplitude from output port $|a^{(2)}_\text{out1(2)}\rangle$ with atoms in state $|xy\rangle$ ($xy=\alpha \alpha, \alpha \beta, \beta \alpha \text{ or } \beta \beta$). $|f^{xy,N}_\text{out1(2)}(t)|$ is the normalized pulse amplitude. The output pulse from $|a^{(2)}_\text{out2}\rangle$ has a $\pi$ phase shift with respect to the input pulse, and this is not shown in (b). In these two figures the input pulse length $T=400\kappa$ and pulse width $w=T/5$ are adopted. The CQED parameters are $g=3\kappa$, $\kappa=\gamma$, and $\kappa_1=\kappa_2=0.45\kappa$. }
\end{figure}

\begin{figure}
\includegraphics[width=8.5cm]{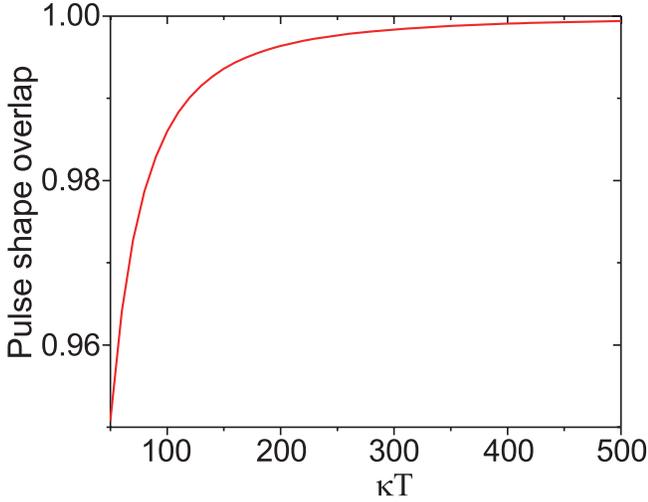}
\caption{\label{fig4} (Color online) Overlap between two normalized output pulse shapes from paths 1 and 2 with the two atoms being in state $|\alpha,\alpha\rangle$ and $|\beta,\beta\rangle$ versus input pulse length $T$. The input pulse width $w=T/5$ and CQED parameters with $g=3\kappa$, $\kappa=\gamma$ are adopted.}
\end{figure}

By using Eqs.~(\ref{eq21}) and (\ref{eq22}) the output pulse shapes in two output modes $|a^{(2)}_\text{out1}\rangle$ and $|a^{(2)}_\text{out2}\rangle$ with single-photon pulses incident in $|a^{(1)}_\text{in1}\rangle$ of interferometer in FIG.~\ref{fig2}(a) can be exactly evaluated under different state combinations of two coupled atoms. Figure ~\ref{fig3} shows the output photon pulse shape from these two output modes with a Gaussian-shaped single-photon pulse as the input under different state combinations of the two atoms. There are two features for the output pulses:

(1) Because of the slow response of the cavity to the input single photon pulse, the output pulse with the atom in state $|\alpha\rangle$ (empty atom) has a shape mismatch to the pulse shape with the atom in state $|\beta\rangle$, where the input pulse is directly reflected. The pulse mismatch will make the two paths to one detector distinguishable and cause deterioration of the generated entangled state. Especially for paths 1 and 2 in FIG.~\ref{fig2}(a) the single photon is reflected or transmitted twice, thus the mismatch between the output pulses is biggest. For paths 3 and 4, both of them evenly experience transmission and reflection once thus have no mismatch between them. The slower variation of the input pulse shape $f_\text{in}(t)$, the fewer mismatches between the two output pulses. A plot of the overlap between two normalized photon pulse shapes from paths 1 and 2 with the two atoms being in states $|\alpha,\alpha\rangle$ and $|\beta,\beta\rangle$ versus input pulse length $T$ is given in Fig.~\ref{fig4}. We can see the overlap is greater than 0.999 when a long enough input pulse is adopted.

(2) Due to the unexpected losses ${{\kappa }_\text{loss}}$, $\gamma$, and limited coupling strength $g$, the coefficients for transmission and reflection are smaller than 1 whenever the atom is in the state $|\alpha\rangle$ or $|\beta\rangle$. Thus the generated states associated with clicks on D1 or D2 are not maximally entangled. There are other states mixed into due to the imperfect transmission and reflection.

With a single-photon pulse injected into the system and D1 clicks, the atomic state can be written as
\begin{equation}\label{eq23}
\begin{aligned}
|\Psi^1_{21}\rangle= \frac{1}{2\sqrt{P^1_{21}}}(A1 |\alpha \alpha\rangle+B1 |\beta \beta\rangle +C1 |\alpha \beta\rangle+D1 |\beta \alpha\rangle ),
\end{aligned}
\end{equation}
with coefficients $A1=\sqrt{\int^T_0 |f^{\alpha \alpha}_\text{out1}(t)|^2dt}$, $B1=\sqrt{\int^T_0 |f^{\beta \beta}_\text{out1}(t)|^2dt}$, $C1=\sqrt{\int^T_0 |f^{\alpha \beta}_\text{out1}(t)|^2dt}$, and $D1=\sqrt{\int^T_0 |f^{\beta \alpha}_\text{out1}(t)|^2dt}$. Here we assume that $\varphi_1=\varphi_2=0$ and the phase difference between the two paths is well controlled, so that $\Delta \varphi=0$. $P^1_{21}=(A1^2+B1^2+C1^2+D1^2)/4$ is the probability of detecting the input photon by D1. The fidelity of this state to the maximum entangled state $|\Phi\rangle=(|\alpha \alpha\rangle+|\beta \beta\rangle)/\sqrt{2}$ is then
\begin{equation}\label{eq24}
F^1_{21}=\sqrt{\frac{(A1+B1)^2}{2(A1^2+B1^2+C1^2+D1^2)}}.
\end{equation}

Through a similar process we can also get the atomic state after D2 clicks, it is
\begin{equation}\label{eq25}
\begin{aligned}
|\Psi^1_{22}\rangle= \frac{1}{2\sqrt{P^1_{22}}}(A2 |\alpha \alpha\rangle+B2 |\beta \beta\rangle +C2 |\alpha \beta\rangle+D2 |\beta \alpha\rangle )
\end{aligned}
\end{equation}
with coefficients $A2=\sqrt{\int^T_0 |f^{\alpha \alpha}_\text{out2}(t)|^2dt}$, $B2=\sqrt{\int^T_0 |f^{\beta \beta}_\text{out2}(t)|^2dt}$, $C2=\sqrt{\int^T_0 |f^{\alpha \beta}_\text{out2}(t)|^2dt}$, and $D2=\sqrt{\int^T_0 |f^{\beta \alpha}_\text{out2}(t)|^2dt}$. $P^1_{22}=(A2^2+B2^2+C2^2+D2^2)/4$ is the probability of detecting the input photon by D2. The fidelity of Eq.~(\ref{eq25}) to the maximum entangled state $|\Psi\rangle=(|\alpha \beta\rangle+|\beta \alpha\rangle)/\sqrt{2}$ is
\begin{equation}\label{eq26}
F^1_{22}=\sqrt{\frac{(C2+D2)^2}{2(A2^2+B2^2+C2^2+D2^2)}}.
\end{equation}

For a CQED system with achievable parameters such as $g=2 \kappa$ or $g=3 \kappa$, $\kappa=\gamma$, and $\kappa_\text{loss}=0.1 \kappa$, from Eqs.~(\ref{eq24}) and (\ref{eq26}) the single-photon detection by D1 and D2 already gives a fidelity of corresponding generated states about 0.98. They are less than unity except when $A1=B1, C2=D2$, and $C1=D1=0, A2=B2=0$ in the ideal case with ${{\kappa }_\text{loss}}=0$ and $g \gg (\kappa, \gamma)$. However, as long as $C1$ and $D1$ ($A2$ and $B2$) can be controlled smaller than $A1$ and $B1$ ($C2$ and $D2$), which is easy to achieve in a low extra loss cavity, the fidelities can be further improved by simply sending more photons into the system and detecting them individually on the corresponding output mode. If we send $n$ single photons into the system one by one and after the $n$th click of D1 or D2, the wave function of the two atoms collapses into
\begin{equation}\label{eq27}
\begin{aligned}
|\Psi^n_{21}\rangle= & \frac{1}{2 \sqrt{P^n_{21}}}(A1^n |\alpha \alpha\rangle+B1^n |\beta \beta\rangle \\
 &+C1^n |\alpha \beta\rangle+D1^n |\beta \alpha\rangle )
\end{aligned}
\end{equation}
or
\begin{equation}\label{eq28}
\begin{aligned}
|\Psi^n_{22}\rangle= & \frac{1}{2 \sqrt{P^n_{22}}}(A2^n |\alpha \alpha\rangle+B2^n |\beta \beta\rangle \\
 &+C2^n |\alpha \beta\rangle+D2^n |\beta \alpha\rangle),
\end{aligned}
\end{equation}
where
\begin{equation}\label{eq29}
 P^{n}_{21}=(A1^{2n}+B1^{2n}+C1^{2n}+D1^{2n})/4
\end{equation}
and
\begin{equation}\label{eq30}
 P^{n}_{22}=(A2^{2n}+B2^{2n}+C2^{2n}+D2^{2n})/4
\end{equation}
are the probabilities of detecting the $n$th photon after $n-1$ photons have been detected by the same detector D1 or D2. As such, the generated entangled states Eqs.~(\ref{eq27}) and (\ref{eq28}) have fidelities
\begin{equation}\label{eq31}
F^n_{21}=\sqrt{\frac{(A1^n+B1^n)^2}{2(A1^{2n}+B1^{2n}+C1^{2n}+D1^{2n})}}
\end{equation}
and
\begin{equation}\label{eq32}
F^n_{22}=\sqrt{\frac{(C2^n+D2^n)^2}{2(A2^{2n}+B2^{2n}+C2^{2n}+D2^{2n})}}.
\end{equation}

\begin{figure}
\includegraphics[width=8.5cm]{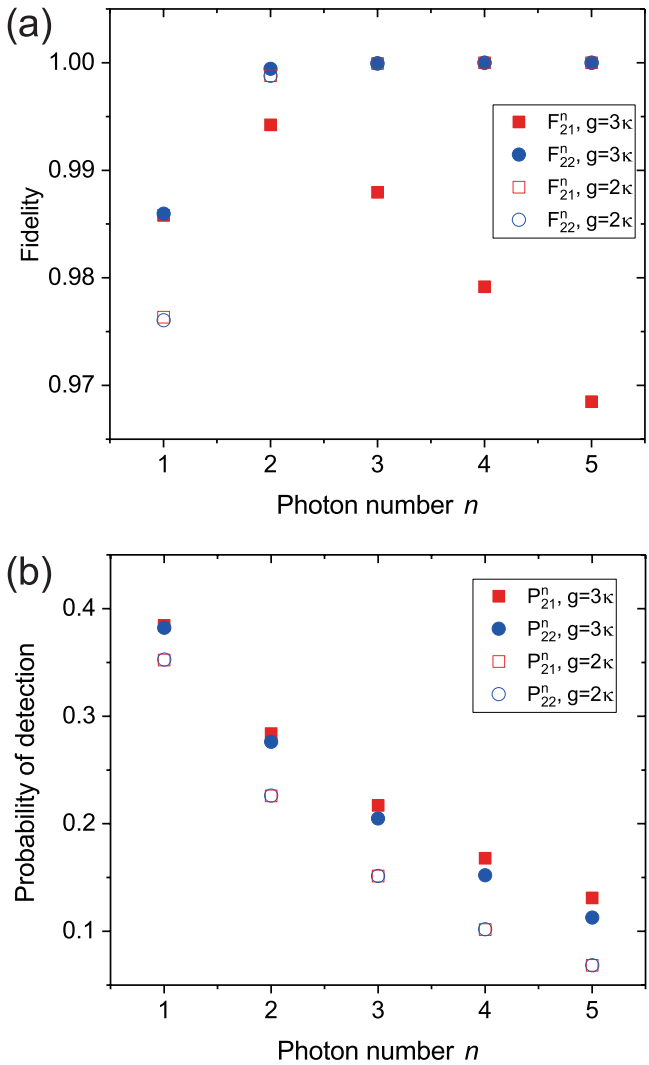}
\caption{\label{fig5} (Color online) Output pulse shapes from two output modes $a^{(2)}_\text{out1}$ (a) and $a^{(2)}_\text{out2}$ (b) with single Gaussian pulse incident from $a^{(1)}_\text{in1}$ port of interferometer in FIG.~\ref{fig2}(a) under different state combinations of two atoms. All the data points in these two figures are calculated by setting the input pulse length $T=400\kappa$ and pulse width $w=T/5$. The CQED parameters are $g=3\kappa$, $\kappa=\gamma$, and $\kappa_1=\kappa_2=0.45\kappa$.}
\end{figure}

The fidelities [Eqs.~(\ref{eq31}) and (\ref{eq32})] versus detected photon numbers are plotted in Fig.~\ref{fig5}(a). We can see that the detection of two photons does enhance the two fidelities to be greater than 0.99 in both cases of $g=2 \kappa$ and $g=3 \kappa$. However, in the case of $g=3 \kappa$ the fidelity shown in Eq.~(\ref{eq31}) decreases when more photons are detected. This is because that the coefficients of transmission for an empty cavity (atom in $|\alpha\rangle$) and reflection for a strongly coupling CQED (atom in $|\beta\rangle$) are not the same in a typical system, and the photon that follows path of 1 or 2 in FIG. ~\ref{fig2}(a) reflects or transmits both of the two cavity systems, which makes the coefficients $A1\neq B1$. When more photons are injected and detected, the difference between $A1^n$ and $B1^n$ becames bigger. Thus, the fidelity shown in Eq.~(\ref{eq31}) decreases. However, for the state given by Eq.~(\ref{eq28}), the photon follows path 3 or 4, where it evenly experiences both transmission and reflection once, so the resulting coefficients $C2$ and $D2$ are the same. The fidelity Eq.~(\ref{eq32}) will approach infinitely to unity with more photons. Anyway, by only two photons the average fidelity can already be substantially enhanced from 0.986 to 0.997 in the case of $g=3 \kappa$. In a special lower coupling example of $g=2 \kappa$, where $A1\approx B1$, both of the fidelities can be further improved. This also provides a method to improve the fidelity through tuning the coupling strength to a suitable value and making $A1\approx B1$ and $C2\approx D2$.  This is feasible in a typical CQED system because the coupling $g$ can be tuned by intentionally moving the relative position of the atom with respect to the cavity mode.

Due to the decay of the atom and unexpected loss of the cavity, the input photon could be decayed out of the system and not be detected. However, once the input photons are detected by the corresponding detectors the combined atomic state collapses to the entangled states. The probability of photon detection is also the probability of successfully generating the entangled states. In the example of $g=3\kappa$ the detection of one input photon on both detectors is about 0.38, which means a total success probability of 0.76. We have already shown that higher fidelity can be achieved by more photons. Figure~\ref{fig5}(b) shows the variation of success probability versus the detected photon number. By using more photons the fidelity of the entangled state can indeed be enhanced at a cost of low success probability. In the example of $g=3\kappa$ the probabilities of two photon detection on both detectors are around 0.28. This means that with the local operations on either one of the atoms the entangled state shown in Eq.~(\ref{eq12}) or (\ref{eq13}) can be prepared with fidelity of 0.997 and success probability of 0.56. With suitable coupling strength and by using more photons the fidelity can be further improved, but the success probability also drops farther. For example, in the example with coupling strength $g=2\kappa$, the fidelity can be further enhanced to over 0.999 by three photons, but the success probability drops to 0.3 (0.15 for each state).

\section{Conclusion and discussion}
In conclusion, we have presented a scheme to entangle multiple remote material qubits through single photons via the nonexcitation process in strongly coupling systems. The basic idea is based on the material state dependent reflection and transmission for the input photons. If two of the strongly coupled systems are arranged as a Mach-Zehnder-interferometer-like configuration, the indistinguishability of the photon paths will finally result in the entanglement of the material qubits in the strongly coupling systems. The entangled state fidelity and success probability are analyzed in detail through strict Heisenberg equations when single photons are injected in the pulsed mode. Our analysis shows that by adopting current achievable system with $g=3\kappa$ the expected entangled state has a fidelity of about 0.986 to the maximum entangled state and success probability about 0.76 with only one photon. If two photons are used, the fidelity could be improved to over 0.99 at the cost of a lower success probability of 0.56. Using a suitable coupling strength the fidelity can be improved further with more photons and lower success probability.

Moreover, the character of no excitation of material qubits guarantees continuity and coherence of material qubits throughout the whole interaction process. Thus our scheme can be directly scaled up to entangle more qubits. Two possible configuration of entangling three or more qubits are also briefly discussed.

In our paper we discussed the fidelity and success probability in the frame of currently accessible optical FP cavity based CQED systems, but our scheme is not only executable on this system. The rapid development of fabrication of micro- or nanostructures and new materials provides more and more new strongly coupled systems \cite{Dayan08, Aoki09, Scheucher2016, Shomroni14, Tiecke14, Goban15, Kato15, Oshea13, Johansson06, Wallraff04, Englund07, Hennessy07, Fink08, Hoi11, Chang07, Tiarks14, Gorniaczyk14, Manzoni14} which can also implement our scheme to entangle the remote material qubits. Especially,  in the strongly coupling system between a single atom and whispering-gallery-mode optical micro-resonator fibers are used to couple the photons in and out of the system \cite{Dayan08, Aoki09, Scheucher2016, Shomroni14}. This provides an easy way to connect the two setups and control the phase difference between different paths as shown in FIG.~\ref{fig2}. So the implementation of our scheme could be more direct.

\begin{acknowledgments}
This work has been supported by the National Key Research and Development Program of China (Grant No. 2017YFA0304502), the National Natural Science Foundation of China (Grants No. 11634008, No. 11674203, No. 11574187, and No. 61227902), and the Fund for Shanxi ¡°1331 Project¡± Key Subjects Construction.
\end{acknowledgments}

\appendix
\section{Transform from Heisenberg picture to Shr\"odinger picture}
The Hamiltonian which governs cavity input-output fields is \cite{Gardiner1985}
\begin{equation} \label{a1}
H_\text{cav,io}= i \int^{\infty}_{-\infty}\text{d} \omega \sum_{i=1,2} \kappa_i(\omega) [a a^\dagger_i(\omega) - a^\dagger a_i(\omega)],
\end{equation}
where $a$ is the annihilation operator for the cavity field, $a_i(\omega)$ is the annihilation operator for outside continuous fields and $\kappa_i(\omega)$ is the input-output coupling efficiency. In the Markov approximation where $\kappa_i(\omega)$ is independent of the frequency $\omega$, the Eqs.~(\ref{eq2}) and (\ref{eq16}) can be deduced from solving the Heisenberg equations \cite{Walls2008}. So the total Hamiltonian should be
\begin{equation} \label{a2}
H_\text{total}= H'+ H_\text{cav,io},
\end{equation}
here $H'$ is shown in Eq.~(\ref{eq17}). Because state $|\beta, 0, \text{vac1},\text{vac2} \rangle$ is the ground state for $H_\text{total}$ with the eigen energy 0, thus $e^{-iH_\text{total} t /\hbar} |\beta, 0, \text{vac1},\text{vac2} \rangle = |\beta, 0, \text{vac1},\text{vac2} \rangle$.

Thus for any operator $A(t)$ we have
\begin{equation} \label{a3}
\begin{split}
 & \langle \beta, 0, \text{vac1}, \text{vac2}|A(t)|\psi(0)\rangle  \\
 &= \langle \beta, 0, \text{vac1}, \text{vac2}|e^{iH_\text{total} t /\hbar}A(0) e^{-iH_\text{total} t /\hbar}|\psi(0)\rangle \\
 &= \langle \beta, 0, \text{vac1}, \text{vac2}|A|\psi(t)\rangle
\end{split}
\end{equation}
and
\begin{equation} \label{a4}
\begin{split}
 & \langle \beta, 0, \text{vac1}, \text{vac2}|{\mathrm d A(t)}/{\mathrm d t}|\psi(0)\rangle  \\
 &=\mathrm d \left[ \langle \beta, 0, \text{vac1}, \text{vac2}|e^{iH_\text{total} t /\hbar}A(0) e^{-iH_\text{total} t /\hbar}|\psi(0)\rangle \right]/\mathrm d t\\
 &= \langle \beta, 0, \text{vac1}, \text{vac2}|\mathrm d \left[A|\psi(t)\rangle\right]/\mathrm d t,
\end{split}
\end{equation}
where $A=A(0)$ and $A(t)$ are the time independent and time dependent operators in Shr\"odinger and Heisenberg pictures. Substituting Eq.~(\ref{eq20}) into Eqs. (\ref{eq19}), (\ref{eq2}), and (\ref{eq3}) and using relations shown in Eqs.~(\ref{a3}) and (\ref{a4}) we finally have the dynamic functions for the coefficients shown in Eq.~(\ref{eq21}).

\bibliography{entanglement}

\end{document}